\documentstyle[12pt]{article}

\input epsf
\makeatletter

\def\newpic#1{}

\voffset= - 0.5cm

\def\hybrid{\topmargin 0pt      \oddsidemargin 0pt
             \headheight 0pt \headsep 0pt

       \textheight 9in         
             \textwidth 6.25in       
             \textheight 9.5in       
             \marginparwidth 0.0in
             \parskip 5pt plus 1pt   \jot = 1.5ex}
\catcode`\@=11
\def\marginnote#1{}

\newcount\hour
\newcount\minute
\newtoks\amorpm
\hour=\time\divide\hour by60
\minute=\time{\multiply\hour by60 \global\advance\minute by-\hour}
\edef\standardtime{{\ifnum\hour<12 \global\amorpm={am}%
             \else\global\amorpm={pm}\advance\hour by-12 \fi
             \ifnum\hour=0 \hour=12 \fi
             \number\hour:\ifnum\minute<10 0\fi\number\minute\the\amorpm}}
\edef\militarytime{\number\hour:\ifnum\minute<10 0\fi\number\minute}

\def\draftlabel#1{{\@bsphack\if@filesw {\let\thepage\relax
        \xdef\@gtempa{\write\@auxout{\string
           \newlabel{#1}{{\@currentlabel}{\thepage}}}}}\@gtempa
        \if@nobreak \ifvmode\nobreak\fi\fi\fi\@esphack}
             \gdef\@eqnlabel{#1}}
\def\@eqnlabel{}
\def\@vacuum{}
\def\draftmarginnote#1{\marginpar{\raggedright\scriptsize\tt#1}}

\def\draftlabel#1{{\@bsphack\if@filesw {\let\thepage\relax
        \xdef\@gtempa{\write\@auxout{\string
           \newlabel{#1}{{\@currentlabel}{\thepage}}}}}\@gtempa
        \if@nobreak \ifvmode\nobreak\fi\fi\fi\@esphack}
             \gdef\@eqnlabel{#1}}
\def\@eqnlabel{}
\def\@vacuum{}
\def\draftmarginnote#1{\marginpar{\raggedright\scriptsize\tt#1}}

\def\draft{\oddsidemargin -.5truein
             \def\@oddfoot{\sl preliminary draft \hfil
             \rm\thepage\hfil\sl\today\quad\militarytime}
             \let\@evenfoot\@oddfoot \overfullrule 3pt
             \let\label=\draftlabel
             \let\marginnote=\draftmarginnote
        \def\@eqnnum{(\theequation)\rlap{\kern\marginparsep\tt\@eqnlabel}%
\global\let\@eqnlabel\@vacuum}  }

\def\numberbysection{\@addtoreset{equation}{section}
             \def\theequation{\thesection.\arabic{equation}}}

\def\underline#1{\relax\ifmmode\@@underline#1\else
             $\@@underline{\hbox{#1}}$\relax\fi}

\def\titlepage{\@restonecolfalse\if@twocolumn\@restonecoltrue\onecolumn
          \else \newpage \fi \thispagestyle{empty}\c@page\z@
             \def\thefootnote{\fnsymbol{footnote}} }

\def\endtitlepage{\if@restonecol\twocolumn \else  \fi
             \def\thefootnote{\arabic{footnote}}
             \setcounter{footnote}{0}}  
\relax

\makeatletter
\newdimen\normalarrayskip              
\newdimen\minarrayskip                 
\normalarrayskip\baselineskip
\minarrayskip\jot
\newif\ifold             \oldtrue            \def\new{\oldfalse}
\def\arraymode{\ifold\relax\else\displaystyle\fi} 
\def\eqnumphantom{\phantom{(\theequation)}}     
\def\@arrayskip{\ifold\baselineskip\z@\lineskip\z@
         \else
         \baselineskip\minarrayskip\lineskip2\minarrayskip\fi}
\def\@arrayclassz{\ifcase \@lastchclass \@acolampacol \or
\@ampacol \or \or \or \@addamp \or
       \@acolampacol \or \@firstampfalse \@acol \fi
\edef\@preamble{\@preamble
      \ifcase \@chnum
         \hfil$\relax\arraymode\@sharp$\hfil
         \or $\relax\arraymode\@sharp$\hfil
         \or \hfil$\relax\arraymode\@sharp$\fi}}
\def\@array[#1]#2{\setbox\@arstrutbox=\hbox{\vrule
         height\arraystretch \ht\strutbox
         depth\arraystretch \dp\strutbox
         width\z@}\@mkpream{#2}\edef\@preamble{\halign
\noexpand\@halignto
\bgroup \tabskip\z@ \@arstrut \@preamble \tabskip\z@ \cr}%
\let\@startpbox\@@startpbox \let\@endpbox\@@endpbox
      \if #1t\vtop \else \if#1b\vbox \else \vcenter \fi\fi
      \bgroup \let\par\relax
      \let\@sharp##\let\protect\relax
      \@arrayskip\@preamble}
\def\eqnarray{\stepcounter{equation}%
                  \let\@currentlabel=\theequation
                  \global\@eqnswtrue
                  \global\@eqcnt\z@
                  \tabskip\@centering
                  \let\\=\@eqncr
     \halign to \displaywidth\bgroup
        \eqnumphantom\@eqnsel\hskip\@centering
        $\displaystyle \tabskip\z@ {##}$%
        \global\@eqcnt\@ne \hskip 2\arraycolsep
             $\displaystyle\arraymode{##}$\hfil
        \global\@eqcnt\tw@ \hskip 2\arraycolsep
             $\displaystyle\tabskip\z@{##}$\hfil
             \tabskip\@centering
        &{##}\tabskip\z@\cr}
\begingroup\ifx\undefined\newsymbol \else\def\input#1 {\endgroup}\fi
\newfont{\hr}{msbm10}
\newfont{\ams}{msam10}

\def\beq{\begin{equation}}
\def\eeq{\end{equation}}
\def\ba{\beq\new\begin{array}{c}}
\def\ea{\end{array}\eeq}

\def\p{\partial}

\def\Doil{{\sf D_{\rm oil}}}
\def\Dalpha{{\sf D}_{\alpha}}

\def\D1{{\sf D}_{1}}

\def\aalpha{{\sf a}_{\alpha}}
\def\balpha{{\sf b}_{\alpha}}
\def\double{{\mit\Sigma}}
\def\curve{{\mit\Gamma}}
\def\pole{p}
\def\qole{q}
\hybrid

\begin{document}

\begin{titlepage}

\title{Whitham hierarchy in growth
problems\footnote{Based on the talk given at
the Workshop ``Classical and quantum integrable systems,
Dubna, January 2004}}

\author{A.~Zabrodin
\thanks{Institute of Biochemical Physics,
4 Kosygina st., 119991, Moscow, Russia
and ITEP, 25 B.Cheremushkinskaya, 117259,
Moscow, Russia}}

\date{March 2004}
\maketitle

\begin{abstract}

We discuss
the recently established equivalence between
the Laplacian growth in the limit of zero
surface tension and the universal Whitham hierarchy
known in soliton theory. This equivalence allows one
to distinguish a class of exact solutions to the
Laplacian growth problem in the multiply-connected case.
These solutions
corerespond to finite-dimensional
reductions of the Whitham hierarchy representable as
equations of hydrodynamic type which are
solvable by means of the generalized hodograph method.

\end{abstract}

\vfill

\end{titlepage}

\section{Introduction}

Dynamics of a moving front (an interface) between two distinct phases
driven by a harmonic scalar field often
appears in different physical and mathematical contexts
and has a number of important practical applications.
Processes  of this type are unified by the name
of Laplacian growth. Their key common feature
is a harmonic field which serves as a potential for the
growth velocity field.

The most extensively studied is the case of
two-dimensional spatial geometry.
To be definite, we shall speak about an interface between two
immiscable and
incompressible fluids with very different viscosities on the plane.
In practice the 2D geometry is
realized in a narrow gap between two parallel
glass plates (the Hele-Shaw cell).
In this version, the problem is also  known
as the Saffman-Taylor problem.
For a review, see \cite{RMP}.

To be more precise, consider a compact
plane domain occupied by
a fluid with a negligible viscosity (water).
We call it water droplet.
Let its exterior be occupied by a viscous fluid (oil).
The oil/water interface is assumed to be a smooth closed
contour.
Oil is sucked out with a constant rate through a sink
(a pump) placed
at some fixed point (which may be at infinity) while
water is injected
into the water droplet.
More generally, one may consider several oil pumps
and several disconnected water droplets.
In this
general setting the problem was discussed in
\cite{Rich}-\cite{EV}.

In the oil domain, the local velocity
$\vec V=(V_x,V_y)$ of the fluid is proportional to
the gradient of pressure $p=p(x,y)$
(Darcy's law):
$$
\vec V=-\kappa \nabla p
$$
where $\kappa$ is called
the filtration coefficient. It is
inversely proportional to viscosity
of the fluid.
For future convenience, we choose units
in such a way that $\kappa$ for oil be
equal to $\frac{1}{4}$.
In particular, the Darcy law holds on the
outer side of the interface thus
governing its dynamics:
$V_n=-\frac{1}{4}\p_n p$.
Here $V_n$ is the growth velocity, which is
by definition normal to the interface, and
$\p_n $ is the normal derivative.
Since the fluids are incompressible ($\nabla \vec V =0$)
the Darcy law implies
that pressure $p$
is a harmonic function in the exterior (oil) domain
except at the point where the oil pump is placed.
There is a logarithmic singularity at this point.

So, the pressure field is a solution
of the time-dependent boundary value problem for the Laplace
operator $\Delta =\p_{x}^{2}+\p_{y}^{2}$
with a prescribed logarithmic singularity and
with certain condition on the boundary
of the growing domain.
This condition is fixed by the following reasoning.
If viscosity of water is small enough,
pressure is uniform inside
water droplets ($p=\mbox{const}$).
Hereafter, we assume that pressure
does not jump across the boundary, so the function
$p(x,y)$ is constant along it.
This means that the surface tension is set to be zero.
To summarize, the mathematical
formulation of the Laplacian growth,
in the limit of zero surface tension, is:
\beq
\label{lg1a}
\begin{array}{l}
V_n=-\frac{1}{4} \p_n p\,,
\\
\Delta p =0 \quad \mbox{in oil with $dp=0$ along the boundary,}
\\
p (x,y)  = \log \rho^{2} +... \quad \mbox{as $\rho \to 0$}
\end{array}
\eeq
Here $\rho$ is the distance between the sink and the point
$x,y$. If the sink is at infinity, what is often
assumed in the outer growth problem,
the last condition
is replaced by
the asymptotics $p = -\log (x^2 +y^2) +\ldots$
very far away from the droplet.
If the oil domain is multiply-connected
(i.e., there are more than one water droplet),
this formulation should be supplemented by
some additional physical conditions for pressure in the water
droplets, which we do not discuss here.
The problem consists in finding the time
evolution of the interface between the fluids.

The simply-connected case (one water droplet)
allows for an effective application of the conformal
mapping technique (see, e.g., \cite{RMP,How}).
Passing to the complex coordinates
$z=x+iy$, $\bar z =x-iy$,
one may describe the growth process in terms of
a time dependent
conformal map $z=f(w,T)$ from a fixed domain
of a simple form,
say the exterior of the unit disk, onto the
evolving oil domain. The interface itself is the image
of the unit circle $w=e^{i\phi}$, $0\leq \phi \leq 2\pi$.
The dynamics (\ref{lg1a}) is then translated to
a nonlinear partial differential equation for
the function $f(w,T)$,
referred to as the Laplacian growth equation.
If the oil sink is at
infinity, it reads
\beq\label{lge}
2{\rm Im}\, \left (\frac{\p f(e^{i\phi},T)}{\p \phi}
\, \overline{\frac{\p f(e^{i\phi}, T)}{\p T}}\right ) =1
\eeq
It first appeared in 1945 in the works \cite{1945}
on the mathematical theory of oil production.

Later,
some exact finite dimensional reductions
of the Laplacian growth equation were found.
Asuming that the derivative
of the conformal map $f$ is a rational function,
it reduces, upon a direct substitution,
to a finite dynamical system
for parameters of this function (say,
for its poles, zeros or residues). See \cite{SB},
\cite{M-WP-D} for
different versions of such dynamical systems.
Moreover, it turnes out that these
systems can be
actually integrated in the sense that
all time derivatives are removable.
As a result, the poles
are represented as implicit functions of time.

Existence of finite-dimensional reductions
was a serious motivation for searching an underlying integrable
structure of the model.
In case of one water droplet,
such a structure
was identified in \cite{M-WWZ} to be the 2D Toda hierarchy in the
limit of zero dispersion.
The Laplacian growth equation
is the string equation of the Toda hierarchy.
The conformal map plays the
role of the Lax function.

The vector field
$\p /\p T$ in the space of domains defines a particular
flow of the Toda hierarchy.
Other flows are ``frozen'' until
oil is sucked by a single pump only.
To ``unfroze'' them, one should allow
oil pumps to work at any point in the plane.
In general there are as many flows as
types and locations of oil pumps.
It is convenient to associate
to each pump its own time variable which
is the total amount of oil sucked by
the pump. It is known \cite{EV} that
the result of the Laplacian growth evolution,
in the case of several pumps,
depends only on the total amount of oil sucked
by each of them and does not depend on the order
in which they operate. In the context of
integrable systems, this fact reflects commutativity
of flows of the Toda hierarchy.

Recently, this approach was extended \cite{LGW}
to the Laplacian
growth of several water droplets. In this case
the dynamics turns out to be equivalent to
the universal Whitham hierarchy on the moduli space
of genus $g$ Riemann surfaces,
where $g$ is the number of water droplets
minus 1.
In full generality, this
hierarchy was introduced by Krichever
\cite{kri1,kri2} in the context of
slow modulations of exact periodic
solutions to soliton equations.
The dispersionless Toda
hierarchy is its simplest particular case at $g=0$.
The universal Whitham hierarchy is known to
have a lot of meaningful finite-dimensional reductions
for any $g$. They are
systems of differential equations of hydrodynamic type which
can be implicitly solved by the
generalized hodograph method \cite{Tsarev}.
Applying these results to the Laplacian growth problem,
we obtain a large family of solutions in the
multiply-connected case. The infinite dimensional
contour dynamics for them is reduced to finite systems
of differential equations which can be integrated.

The aim of the present paper is to
study this class of
finite-dimensional reductions
and to clarify their algebro-geometric
meaning.

\section{Laplacian flows and the universal
Whitham hierarchy}

In this section we explain the precise meaning
of the statement that the Laplacian growth
is described by the Whitham equations.

\begin{figure}[tb]
\epsfysize=4cm
\centerline{\epsfbox{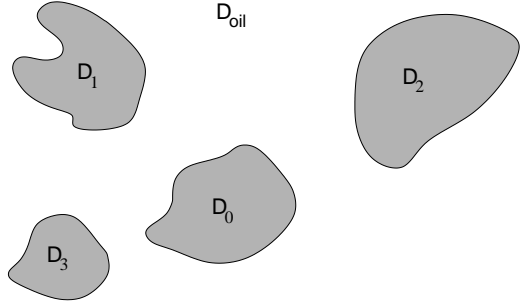}}
\vspace*{0.5cm}
\caption{\sl Water droplets
${\sf D}_{\alpha}$ ($g=3$).}
\label{fi:droplets}
\end{figure}

We use the following notation (Fig.~\ref{fi:droplets}).
Let $\Doil$ be the region of the plane occupied
by oil (a non-compact
domain containing infinity), and ${\sf D}$ be the region occupied by
water.
We assume that there are $g+1$ water droplets,
which are compact domains
bounded by smooth non-intersecting
closed contours $\gamma_{\alpha}$,
$\alpha =0,1, \ldots , g$.
Let $\Dalpha $ be the $\alpha$-th water droplet,
so that ${\sf D}$ is their union. By $\gamma$ we
denote the union of the $\gamma_{\alpha}$'s:
$\gamma =\cup_{\alpha =0}^{g}\gamma_{\alpha}$.
Our assumptions imply that no contour $\gamma_{\alpha}$
lies inside another.

\subsection{Commutativity of
Laplacian flows}

Any growth process defines a flow in the space
of contours\footnote{Depending
on a particular problem,
the flow can be only defined on a proper subvariety
or a subdomain in the space of contours, but this is not
important for us here.}, or, in the infinitesimal version,
a vector field.
Integral curves of this vector field are
just one-parametric families of
contours $\gamma (T)$ obtained from
any initial contour $\gamma (0)$ by the
evolution in the time $T$.

Flows corresponding to the Laplacian growth
(Laplacian flows) are determined by location
of oil pumps. Namely, given an oil sink
at a point $a \in \Doil$, the solution of the
boundary value problem
\beq\label{bp}
\Delta p (z) = 4\pi \delta^{(2)}(z-a)\,,
\quad \quad dp=0 \; \mbox{on $\gamma$}
\eeq
in $\Doil$ (here $\delta^{(2)}(z-a)$ is the two-dimensional
delta-function supported at the point $a$), with
some additional conditions on the boundaries of
$\Dalpha$'s in case of more than one water droplet
(see \cite{LGW} for details), uniquely defines
an infinitesimal deformation of $\gamma$ by setting
the vector of normal velocity to be $-\frac{1}{4}\p_n p$.
Such a deformation
is a tangent vector to the space of contours.
For brevity, in this paper we do not discuss
other types of Laplacian flows considered in \cite{LGW}.

Given a Laplacian flow,
a natural parameter along the trajectories
in the space of contours is the amount of oil, $T$,
sucked by the pump which defines the flow.
Since the both fluids are assumed to be incompressible,
$T$ is simply the change of the total area
of the water droplets. (If the pumping rate is
constant, $T$ can be also identified with time.)
Let $\p_T$ be the corresponding vector field.
It is defined
in an invariant way, i.e., it does not depend on the choice
of local coordinates in the space of contours.

As is common, we identify
vector fields with derivatives along them.
The derivative
of any functional $F$ on the space of contours
along the vector field (the Lie derivative)
is defined as follows:
$$
\p_T F = \lim_{\epsilon \to 0}
\frac{F(\gamma (T+\epsilon )) -F(\gamma (T))}{\epsilon}
$$
If there are several sinks (at the points $a$, $b$,
etc), we write
$\p_{T^{(a)}}$, $\p_{T^{(b)}}$, etc for the corresponding
vector fields.

The notation is justified by the following theorem \cite{EV}.

\noindent
{\bf Theorem 1.}
{\it The vector fields $\p_{T^{(a)}}$, $\p_{T^{(b)}}$
commute for all $a,b \in \Doil$.}

\noindent
It is an infinitesimal formulation of the fact that
the result of the Laplacian growth evolution
(i.e., the shape of $\Doil$) depends
only on the amounts of oil $T^{(a)}$,
$T^{(b)}$ sucked from the points
$a$, $b$ and does not depend on other
details of the process (assuming that
no topological
change or singularity occurs).
In its turn, this fact follows from
the reduction of the
Laplacian growth to the
inverse potential problem in two dimensions
and from local uniqueness of solutions to the latter.
For more details see \cite{EV}.

This result suggests to treat
$\p_{T^{(a)}}$ as a partial derivative
in a space with coordinates $T^{(a)}$.
To construct
a simple example of such a space,
fix $N$ points $a_j$ and consider the variety of
contours which can be obtained from some initial
configuration as a result of all
Laplacian growth processes
with oil pumps at the points $a_j$. The resulting shape
of the droplets is uniquely determined (if no singularity
occurs) by amounts of oil sucked out of each point
$a_j$, i.e., by the quantities $T^{(a_j)}$.
This configuration space is $N$-dimensional and
$T^{(a_j)}$ are local coordinates in it.

\subsection{The Schottky double}

Our next task is to
carry the Laplacian flows over to
compact Riemann surfaces with a real structure.
To do that we need the Schottky double
construction \cite{SS} which
we recall here.

With the domain $\Doil$
(as well as with any plane domain with a boundary)
one canonically associates
its Schottky double
$$
\double =\Doil \cup \gamma \cup {\sf D}_{\rm oil}^{*}
$$
where ${\sf D}_{\rm oil}^{*}$ is another copy of
$\Doil$ (its ``opposite side'') attached to it along
the boundary $\gamma$.
In what follows, we
shall call $\Doil$ and ${\sf D}_{\rm oil}^{*}$
the {\it front side} and the {\it back side}
of the double, respectively.
Formally speaking, a point
$P \in \double$ is a pair $P=(z, \sigma )$, where
$z\in \Doil$ and $\sigma =\pm$ indicates front or back side.
If $z\in \gamma$, the pairs $(z, \pm)$ represent the same point.
There is a natural projection
$\double \rightarrow \Doil$
which sends a point $P=(z, \sigma ) \in \double$ to
the point $z\in \Doil$. In the sequel, we
sometimes write simply $z$ for $(z,+)$ and
$z^{*}$ for $(z,-)$.

By adding a point at infinity for each copy of $\Doil$,
$\double$ becomes a compact surface naturally endowed
with a complex structure. The complex structure on
the front side is the same as in $\Doil$ while
the back side has the opposite complex structure.
In other words, the
holomorphic coordinate is $z$
on $\Doil$ and $\bar z$ on
${\sf D}_{\rm oil}^{*}$. Whence $\double$
is a compact Riemann surface with the antiholomorphic
involution $P \mapsto P^{*}$ which sends
$(z, \sigma )$ to $(z, -\sigma )$.
The two sides of the double are in a sense mirror images
of each other, so we shall call points
connected by the involution mirror-symmetric points.
Boundaries of the droplets are fixed points of the involution.
A schematic view of the Schottky double see in
Fig.~\ref{fi:poles} below.

If the contours $\gamma_{\alpha}$
are smooth, the double is, by construction,
a {\it smooth} Riemann surface of genus $g$. Indeed, each point
of $\double$, including points of the boundary
of $\Doil$, has a disk-like neighborhood.

Note that the set of fixed points of the involution
consists of exactly $g+1$ closed contours, which is
the maximal possible number for surfaces of
genus $g$. Riemann surfaces with antiholomorphic involution
having this property are called real.

In what follows it is implied that a
canonical basis of ${\sf a}$ and ${\sf b}$-cycles
on the double is fixed. This can be done
in many ways. For example, one may take $\aalpha$-cycles to be
boundaries of the droplets
$\Dalpha$ for $\alpha =1, \ldots , g$
and take $\balpha$-cycles to be
some paths connecting the boundary of
the $0$-th droplet with
the $\alpha$-th one on the front side and going back
through the opposite side of the double.
Each choice of the basis of cycles
corresponds to certain physical conditions
in the water droplets which are discussed in detail
in \cite{LGW}. For our purpose here it is enough to
imply that one or another basis is fixed.

Any pair
of meromorphic functions $\varphi ,\tilde \varphi$
(which are
functions on $\Doil$ and on the
complex congugate domain respectively)
such that $\varphi (z)=\tilde \varphi (\bar z)$ on the boundary
defines a meromorphic function on the double.
Namely, this function takes the value $\varphi (z)$
at the point $z$ on the front side and
the value $\tilde \varphi (\bar z)$ at the mirror-symmetric
point $z^{*}$ on the back side.
Similarly, a meromorphic differential on the double
is represented by a pair of meromorphic differentials
$\varphi (z)dz$ and $\tilde \varphi
(\bar z)d\bar z$ such that
$\varphi (z)dz =\tilde \varphi
(\bar z)d\bar z$ along the boundary
contours.

If a meromorphic
differential $d \omega (z)$
on $\Doil$ is purely imaginary along the boundary,
the analytic continuation to the opposite side of the
double is especially simple.
Indeed, in accord with
the Schwarz reflection principle, its
meromorphic extension to the back side is
$-\overline{d \omega (z)}$, so for each pole of such a
globally defined differential there is a mirror pole
on the opposite side. In this case we shall say that
$d\omega$ is a {\it symmetric} differential on $\double$.

\subsection{Abelian differentials on the double}

To each pump we assign an Abelian differential on
$\double$. On the front side it is the differential
$\p_z p dz$, where $p$ is the pressure field generated
by the pump in the viscous fluid. Since
$dp=\p_z p dz +\p_{\bar z}p d\bar z =0$ on the boundary,
$\p_z p dz$ is purely imaginary along the boundary.
So, its analytic continuation to the back side of the double is
$-\p_{\bar z}p d\bar z$. Conversely, given a symmetric
Abelian differential on $\double$,
we place pumps at
its poles on the front side. Strengths of the pumps
are determined by the residues.

Note also that any symmetric Abelian differential $d\Omega$
on $\double$ defines a tangent vector to the space
of contours via setting the normal velocity
at any point of the contour to be
$$
V_n =-\frac{1}{2}\p_n \, {\rm Re}
\int^z d\Omega
$$
If poles of $d\Omega$ are fixed in the $z$-plane,
this defines a vector field corresponding to
a Laplacian flow.

For example, if oil is sucked (with a unit rate)
from a point $a \in \Doil$, then
$$
p(z)=2G(a,z)
$$
where $G$ is the Green function of the Dirichlet
problem in $\Doil$. The Green function
is defined as a solution of the Laplace equation in $\Doil$
with the logarithmic singularity of the form
$G(a,z) \to \log |z-a|$ as $z\to a$ and such that
$dG=0$ on the boundaries.
Therefore, the differential
$dW^{(a)}(z)=\p_z p dz$ has a simple pole at
$z=a$ with the residue 1
and no other singularities in $\Doil$. The analytic
continuation of this differential to the back side
is $-\overline{dW^{(a)}(z)}$ which gives
a simple pole at the mirror point $a^{*}$ with
the residue $-1$. The differential constructed is
a symmetric Abelian differential of the third kind
(a dipole differential):
\beq\label{3dkind}
dW^{(a)}=\left \{
\begin{array}{cc}
2\p_z G(a,z)dz & \mbox{on the front side}
\\ &\\
- 2\p_{\bar z} G(a,z)d\bar z & \mbox{on the back side}
\end{array}
\right.
\eeq

\noindent
{\bf Remark.} For $g\geq 1$ these conditions
do not yet define the differential uniquely.
To fix it, a normalization is needed.
We normalize the differential
$dW^{(a)}$ by the condition that it has
zero ${\sf a}$-periods. Depending on the
choice of cycles, this means certain physical
conditions: for instance,
fixed injecting rates or fixed pressures
in the water droplets. See \cite{LGW} for
details.

\subsection{The Whitham hierarchy}

The Whitham hierarchy structure reveals itself
in evolution of the Abelian differentials under
Laplacian flows $\double \to \double (T)$.

The holomorphic coordinate $z$
on the front side of $\double$
is independent of time by construction.
It defines a connection in the space of
real Riemann surfaces of genus $g$.
Correspondingly, the Abelian
differentials become time dependent:
$dW =dW(z, T)$. At any fixed $z$, $dW(z)$
is a functional on the space of contours,
so one can take derivatives along
Laplacian vector fields.

In \cite{LGW}, it was shown that the Laplacian growth
of $\Doil$, with zero surface tension, is equivalent
to the system of Whitham equations for symmetric
Abelian differentials of the third kind
$d W^{(a)}$ on the Schottky double of $\Doil$.
The main result is the following theorem \cite{LGW}:

\noindent
{\bf Theorem 2.}
{\it For any pair of pumps (at the points
$a,b \in \Doil$),
the following relations hold:
\beq\label{W1}
\p_{T^{(a)}}\, dW^{(b)}=
\p_{T^{(b)}}\, dW^{(a)}
\eeq
}

\noindent
This statement follows from
(\ref{3dkind}) and the Hadamard
variational formula \cite{Hadamard} for the
Green function $G(z,a)$. The Hadamard
formula implies that
$\p_{T^{(a)}} G(b,c)$ is symmetric under
permutations of the points $a,b,c$.

The theorem allows one to identify
Laplacian flows with Whitham flows on the
(extended) moduli space of real Riemann surfaces.
Equations (\ref{W1}) constitute
the universal Whitham hierarchy \cite{kri2}.
In the soliton
theory, they are obtained by averaging solutions
of soliton equations over fast oscillations \cite{Whitham}.

Since the Laplacian vector fields commute,
one may look for a general solution to
equations (\ref{W1}) in the form
\beq\label{W2}
dW^{(a)}=-\p_{T^{(a)}}d\Omega
\eeq
where $d\Omega$ is called
{\it the generating differential}.

\noindent
{\bf Remark.}
In general, singularities of the generating
differential may be of a very
complicated nature. Equations (\ref{W2}) mean,
however, that all of them do not move
under the Laplacian flows. In other words, they are
integrals of motion. The only parameter that
changes in the course of the Laplacian growth
with a sink at the point $a$ is the residue
of $d\Omega$ at the pole at $a$.
This means, in particular, that the Laplacian
flows with point-like pumps can only add simple
poles to singularities of $d\Omega$.

In the next section we consider an important
special case when all singularities of
the generating differential are poles,
i.e., $d\Omega$ is meromorphic on $\double$.

\section{Algebraic domains}

Families of special solutions to the Whitham equations
with a meromorphic generating differential
were constructed in \cite{kri1}.
In \cite{LGW}, it has been
shown that they describe Laplacian growth
of algebraic domains.
It is clear that an algebraic domain remains algebraic
under Laplacian flows with point-like pumps.
This fact allows one to reduce
the infinite dimensional configuration space of contours
to a finite dimensional one.

\subsection{The Schwarz function}

For a more explicit characterization
of the class of algebraic domains
that we are going to deal with, consider
the analytic continuation of the
function $z$ defined in $\Doil$ to the back side
of the Shottky double.

\noindent
{\bf Definition.}
{\it The domain $\Doil$ is said to be
algebraic if the function $z$ is extendable to a meromorphic
function on the Schottky double of $\Doil$.}

\noindent
To clarify the meaning of this definiton, we recall
the notion of the Schwarz function \cite{Davis}.
Given a bounded plane domain, the Schwarz function is
the analytic continuation
of the function $\bar z$ away from the boundary.
Note that for such an analytic continuation
to exist, the boundary needs to be analytic, not only
smooth.
By definition,
the Schwarz function, $S(z)$, is a function analytic in some
neighborhood of the boundary contour (contours) such that
$$
S(z)=\bar z
\;\;\;\;
\mbox{on the contour}
$$
If it exists, the Schwarz function must obey the
``unitarity condition''
$$
S(\overline{S(z)})=\bar z
$$
which means that the function $\overline{S(\bar z)}$ is
inverse to $S(z)$.
In general, the analytic continuations from different
components of the boundary of a multiply-connected domain
may lead to different functions.
Algebraic domains correspond to a rather special
situation when they
lead to a common function which is well-defined
everywhere in $\Doil$.

If the Schwarz function exists, the analytic
continuation of the function $z$ to the back side
is just $\overline{S(z)}$.
Therefore,
a plane domain is algebraic if and only if
the Schwarz function
is extendable to a meromorphic function in it.

\begin{figure}[tb]
\epsfysize=6cm
\centerline{\epsfbox{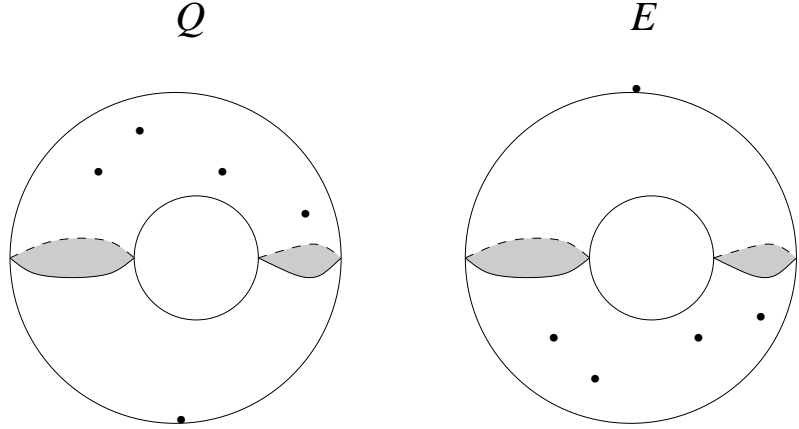}}
\vspace*{0.5cm}
\caption{\sl Poles of the functions $Q$ and $E$ on the
Schottky double ($g=1$). The upper and lower halves
of the torus give a schematic view of
the front and back sides
of the double, respectively.}
\label{fi:poles}
\end{figure}

Let $P\in \double$
be a point on the Schottky double
of an algebraic domain. Consider the function $Q(P)$
on $\double$ defined as follows:
$$
Q(P)=\left \{ \begin{array}{cc}
S(z) &\;\;\mbox{if $P$ is on the front side}
\\ &\\
\bar z &\;\;\mbox{if $P$ is on the back side}
\end{array} \right.
$$
Here $z=z(P)$ is the $z$-coordinate of the point
$P=(z, \sigma )$ on $\double$.
By definition
of the Schwarz function, $Q(P)$ extends continuously
across the contour $\gamma$ and so it is a well-defined
meromorphic function on $\double$. Along with the $Q(P)$,
we introduce a ``mirror-symmetric'' function
$$
E(P)=\left \{ \begin{array}{cc}
z &\;\;\mbox{if $P$ is on the front side}
\\ &\\
\overline{S(z)} &\;\;\mbox{if $P$ is on the back side}
\end{array} \right.
$$
which is meromorphic on $\double$ as well.
It provides the above mentioned
meromorphic extension of the function
$z$ to the whole double.
Note that the functions $Q$, $E$ are connected by the relation
\beq\label{SEQ}
S(E(P))=Q(P)
\eeq
and transform one into another (with complex
conjugation)
under the antiholomorphic involution:
\beq\label{invol}
Q(P^{*})=\overline{E(P)}
\eeq

Suppose that the Schwarz function in $\Doil$
has poles at some points $\pole_j \in \Doil$ with
multiplicities $n_j$.
Hereafter, the points $\pole_j$ are assumed to be
in general position.
Set $N=\sum_j n_j$.
Then the number of droplets can not exceed $N+1$,
or, what is the same, genus of the Schottky double is
less or equal to $N$:
$$
g\leq N= \sum_j n_j
$$
This rough estimate\footnote{This estimate
is rather rough indeed. It
can be improved by
taking into account the reality
conditions. For example, in the case $N=1$ our
inequality gives $g\leq 1$ while
a more detailed analysis (see, e.g.,
Theorem 7 and Corollary 7.1 from \cite{gus})
shows that $g=0$.}
follows from the fact that
any nontrivial function on a Riemann surface of genus $g$
has at least $g+1$ poles counted with multiplicities
(if everything is in general position).
Applying this statement to the function $Q$, we
get the estimate.

To summarize, on the Schottky double of an algebraic
domain there exist two meromorphic functions
connected by the involution as in (\ref{invol}).
The notation $Q$, $E$ is chosen to emphasize
the similarity with the approach developed in
\cite{kri2}.

\noindent
{\bf Remark.}
The class of algebraic domains introduced here
and in \cite{EV}
coincides with the class of ``quadrature domains''
from \cite{gus}.
The class of domains
corresponding to ``algebraic orbits'' of the Whitham
equations \cite{kri2} is broader. It includes
domains for which the generating differential is meromorphic
on $\double$ with ``jumps'' across certain cycles.

\subsection{A complex curve associated with an
algebraic domain}

For algebraic domains, the Schwarz function
is an {\it algebraic function}, i.e.,
it satisfies a polynomial equation
$R(S(z), z)=0$.
The easiest way to see this is to consider
the analytical continuations
of the functions $z$, $S(z)$ to the
whole double, i.e., the functions $E,Q$.

If $S(z)$ has $N$ poles in $\Doil$
(counting multiplicities), then the function $Q$ has
$N+1$ poles on $\double$ ($N$ poles on the front side
and a simple pole at infinity
on the back side). Similarly, the function $E$
also has $N+1$ poles. Therefore, we have two meromorphic
functions with $N+1$ poles on a smooth Riemann surface.
A simple theorem of algebraic geometry states that
these functions are connected by a polynomial equation
of degree $N+1$ in each variable:
\beq\label{RQE}
R(Q,E)=\sum_{n,m =0}^{N+1}A_{nm}Q^n E^m =0
\eeq
Symmetry under the involution implies that the
matrix of coefficients is Hermitian:
$A_{nm} = \overline{A_{mn}}$. Restricting the relation between
$Q$ and $E$
to the front side of the double, we get the polynomial
equation
$$
R(S(z), z)=0
$$
The unitarity condition for the Schwarz function is equivalent
to the hermiticity of coefficients of this polynomial.

The equation $R(\tilde z, z)=0$ defines a complex
algebraic curve $\curve$
which may serve as another realization of the Riemann
surface $\double$. The boundary contour is recovered
as the set of real solutions $x,y$ to the equation
$R(x-iy, \, x+iy)=0$. It is the real
section of the curve $\curve$.

\noindent
{\bf Remark.}
We see that
the boundary of any algebraic domain is always described
by a polynomial equation.
The converse is not true. An arbitrary
polynomial equation with hermitian coefficients
in general does not define an algebraic domain.

Despite of the fact that $\double$ is a
smooth surface,
the curve $\curve$ is not smooth for $N>0$.
It contains singular points (double points
in case of general position). This is seen from
comparing genus of the double with the upper
bound of genus of the curve $\curve$. For a generic
{\it smooth} curve of the form (\ref{RQE}) genus
equals $N^2$, so the fact that the genus does not actually
exceed $N$ just
gives evidence for the presence of singular points
on the curve $\curve$.

\begin{figure}[tb]
\epsfysize=5cm
\centerline{\epsfbox{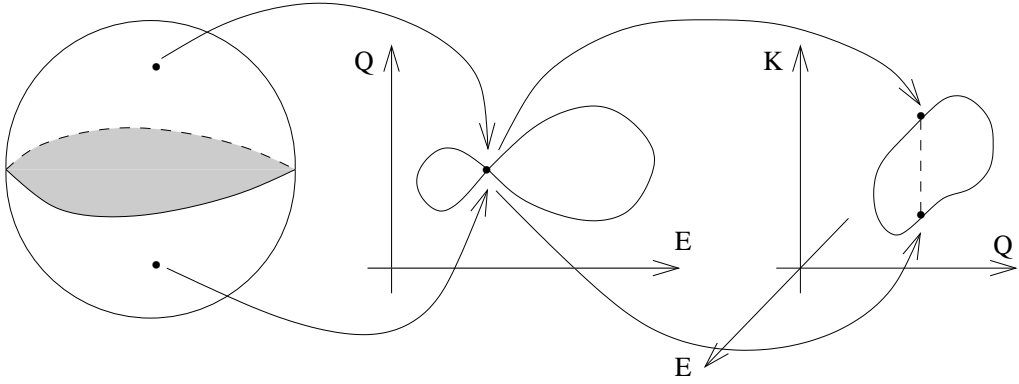}}
\vspace*{0.5cm}
\caption{\sl From a smooth Schottky double
to a singular curve $\curve$ and then
to the smooth curve $\tilde \curve$. }
\label{fi:sing}
\end{figure}

How do the singular points emerge?
Consider the map
$P\mapsto (Q(P), E(P))$ from $\double$ to $\curve$
which takes a point $P$ to the pair of complex numbers
$(Q(P), E(P))$ connected by the polynomial equation.
If two different points of $\double$ are mapped to
one point of $\curve$ (i.e. values of the
both functions $Q,E$ are the same at these points),
then this point is clearly a singular point of the curve,
namely, a self-crossing (or double) point.
This situation is illustrated in
Fig.~\ref{fi:sing}.
In short, we say that the map from the Schottky
double to the curve glues some points together.
This means that $\double$ in general can not
be realized as a smooth algebraic curve in ${\bf C}^2$.

Let us give a more precise
characterization of the points that can be glued together.
Two distinct points $P_1 , P_2 \in \double$ are glued together
if $Q(P_1)=Q(P_2)$ and $E(P_1)=E(P_2)$. As is readily seen, this is
possible only if $P_1$ and $P_2$ are on different sides of the double.
Let $P_1 =(z_1 , +)$, $P_2 =(z_2 , -)$ with $z_{1,2}\in \Doil$.
Then these points are glued together if $S(z_1)=\bar z_2$ and
$S(z_2)=\bar z_1$.
It might seem that the second equality follows from the first one
by virtue of the unitarity condition.
However, the unitarity condition holds
for the whole algebraic function $S(z)$ while
$S(z)$ in the both equalities above means one and the same
branch of the Schwarz function obtained as
the analytic continuation of $\bar z$ to $\Doil$.
So, we have two independent
conditions to determine the points $z_{1,2}$.
To put it differently,
$z_1$ (or $z_2$) is a point in $\Doil$ where two (or more)
branches of the Schwarz function take same value.

Nevertheless, the (smooth) Schottky double always has a realization
as a smooth complex curve $\tilde \curve$
in a higher dimensional space.
This curve can be obtained from $\curve$ as a result
of the desingularization process. In the case of
a self-crossing point this simply amounts to
incorporating a third meromorphic function on
$\double$, say $K(P)$, such that it takes different
values at the points to be glued together by the map
$P\mapsto (Q(P), E(P))$ (see Fig.~\ref{fi:sing}).
The curve
is then defined by a system of polynomial
equations $R_1(Q,K)=0$, $R_2 (K,E)=0$ in ${\bf C}^3$.
If this curve is still singular, the process should be
repeated until one arrives at a smooth curve
$\tilde \curve$.

The so obtained smooth curve $\tilde \curve$
is isomorphic to
the Riemann surface of the Schwarz function.
The latter can be also realized as a branched covering
of the $z$-plave.
In this realization, the antiholomorphic
involution is the Schwarz reflection
$$
z\mapsto \overline{S(z)}
$$
which sends a point $z$ to its ``mirror'' with respect to
the boundary contour\footnote{This geometric interpretation requires
some care.
If the point $z$ is not far from the contour, so
the image belongs to the same sheet of the covering,
this is indeed a kind of reflection in the usual sense.
However, if $z$ is far enough from the contour,
the image goes to another sheet across a branch cut.
In this case the
unitarity condition (involutivity of the
reflection) is fulfilled only after taking
proper branches of the Schwarz function.}.

\subsection{The generating differential}

The explicit form of the generating
differential for algebraic domains is:
$$
d\Omega =QdE =\left \{
\begin{array}{l}
S(z)dz \quad \mbox{on the front side}
\\ \\
\bar z d\overline{S(z)} \quad \mbox{on the back side}
\end{array}
\right.
$$
To prove this, one should check equations (\ref{W2})
for all vector fields $\p_{T^{(a)}}$.
These equations immediately follow from
the behaviour of singularities of the Schwarz
function under Laplacian flows.

\noindent
{\bf Proposition.} {\it The location of poles of the
Schwarz function of any algebraic domain remains
unchanged under Laplacian flows with point-like pumps.
The residues remain unchanged as well, except for the residue
at the pole where the pump is placed:
\beq\label{omega3}
\p_{T^{(b)}}\, {\rm res}\,_{z=a}(S(z)dz)=
\left \{
\begin{array}{c} 0 \;\;\;\;\; \mbox{if} \quad b\neq a\\
-1 \;\;\; \mbox{if} \quad b = a
\end{array}
\right.
\eeq
}

\noindent
{\bf Remark.}
If $S(z)$ is initially regular at a point $a$, the Laplacian
flow with the sink at $a$ generates a simple pole
at this point with the residue $-T^{(a)}$.

\noindent
One of the methods to prove this proposition is
to consider the function
$$
C(z)=\oint_{\gamma}
\frac{\bar \xi d\xi}{z-\xi}
$$
whose variation under Laplacian flows can be found
by a direct calculation. On the other hand,
as it follows from the properties of Cauchy
integrals, singularities
of the Schwarz function outside $\gamma$ (i.e. in $\Doil$)
coincide with singularities of the analytic continuation
of the $C(z)$ from inside to outside $\gamma$.

It follows from the proposition that the
only singularity of the differential
$\p_{T^{(a)}}(QdE)$ on the front side is the simple pole
at $a$ with residue $-1$.
The unitarity condition means that this differential
is purely imaginary on $\gamma$.
Indeed, acting by $\p_{\bar z}$ and
$\p_{T}$ on the unitarity condition
$S\, (\overline{S(z, T)},\,  T)=\bar z$, we
get the identities
\beq\label{id1}
S'(\overline{S(z)})\,\, \overline{S'(z)}=1
\eeq
\beq\label{id2}
\p_T S(\overline{S(z)},\, T)+
S'(\overline{S(z)})\, \p_T \overline{S(z, T)}=0
\eeq
(here $\p_T$ means any of $\p_{T^{(a)}}$).
Combining them, we obtain:
$$
\p_T \overline{S(z,T)dz}=
-\p_T S(\overline{S(z)}, T)d \overline{S(z)}
$$
which yields
$\overline{\p_T Sdz}=-\p_T Sdz$ on $\gamma$.
Taking into account the normalization condition,
we conclude that
\beq\label{W2a}
\p_{T^{(a)}}(QdE)=-dW^{(a)}
\;\;\;\;
\mbox{for all $a\in \Doil$}
\eeq

Let us fix $a$ and set $T=T^{(a)}$, $W=W^{(a)}$.
If one takes the function $W=W(z, T)$ as a connection
in the space of real Riemann  surfaces, instead of
the function $z$,
i.e. keeps $W$ constant when applying the vector
field $\p_T$,
then equation (\ref{W2a})
acquires the form
\beq\label{lge1}
\p_{T} E(W,T) dQ - \p_{T} Q(W,T) dE =dW
\eeq
In the simply-connected case, it turns,
after the restriction to $\gamma$ and
the identification $E=f$, $Q=\bar f$,
$iW=\phi$, into
the Laplacian growth equation (\ref{lge}).

\section{Whitham equations for
Laplacian growth of algebraic domains}

In this section we represent the Whitham
hierarchy (\ref{W1}) as a finite system of
differential equations for branch points of
the Schwarz function and give a general form
of the generalized hodograph solution of this system.

\subsection{Branch points of the Schwarz function}

Given an algebraic domain, consider zeros of
the differential $dE$ on the Schottky double.
Since $E=z$ on the front side, all these zeros
are on the back side. Let us denote them by
$\lambda_{k}^{*}=(\lambda_k , -)$, so
$\lambda_k$'s are their projections on
the $z$-plane. Since $0=dE(\lambda_{k}^{*})=
d\overline{S(\lambda_k)}$, $\lambda_k$'s
are zeros of the derivative
of the Schwarz function in $\Doil$:
$$
S'(\lambda_k)=0\,,
\;\;\;\; \lambda_k \in \Doil
$$

Given poles of the Schwarz function
and the number of droplets, the number of
such points can be found from the following
reasoning.
The differential $dE$ has poles of orders $n_j +1$
at the points $(p_j , -)$ on the back side of the double
(recall that $p_j$ are poles of the Schwarz function)
and the second order pole at infinity on the front side.
The total number of poles is
$2+ \sum_j (n_j +1)$, counting multiplicities.
For any meromorphic
differential on a smooth genus-$g$ Riemann surface
the difference between the numbers of its zeros and poles
equals $2g-2$. Therefore, the number $M$ of zeros
of $dE$ is given by
\beq\label{M}
M=2g +\sum_j (n_j +1)
\eeq
In particular, if all poles $\pole_j$ of the Schwarz function
are simple ($n_j =1$), then
\beq\label{Ma}
M=2(g+N)
\eeq

In general position zeros
of the differential $dE$ are all simple.
Assuming this,
the critical values of the function $E$, i.e.,
values of $E$ at the zeros of $dE$,
\beq\label{crval}
E_k := E(\lambda_{k}^{*})=
\overline{S(\lambda_k )}
\eeq
play the role of local coordinates in the space of algebraic domains
with given $N$ and $g$ (see \cite{kri1}).
They are images of the points $\lambda_k$ under
the Schwarz reflection.
Their geometric meaning is clarified by expanding
the unitarity condition in a vicinity of the points
$\lambda_k$ (or $E_k$).
As a result, we obtain:
\beq\label{branch}
S(z)=S(E_k)+\alpha_k \sqrt{z-E_k}+\ldots \,\, ,
\;\;\;\;\; z\to E_k
\eeq
Therefore, $E_k$'s are branch points of the Schwarz
function. In general position they are of order two.
If all poles of the Schwarz function are
simple, then all the
$2(g+N)$ branch points (see (\ref{Ma})) are inside
water droplets. In this case a single-valued
branch of the Schwarz function can be defined
by making $g+N$ cuts between them.

\subsection{Dynamics of the branch points}

Here we derive differential equations
for dynamics of the branch points of the
Schwarz function $E_i =E_i (T)$.

Let us expand the equality
(\ref{W1})
on $\double$ at
the points where $E$ fails to be a good local
parameter, i.e., at zeros $\lambda_{k}^{*}$
of the differential $dE$.
Near these points the local parameters are
$\sqrt{E-E_k}$, so the expansions have the form:
$$
W^{(a)}=W^{(a)}(\lambda_{k}^{*})+ C^{(a)}_k
\sqrt{E-E_k} + O(E-E_k)
$$
$$
\p_{T^{(b)}}\,d W^{(a)} =
\frac{C^{(a)}_k \p_{T^{(b)}} E_k \,
dE}{2(E-E_k)^{3/2}} \, +
O\left ( \frac{dE}{\sqrt{E-E_k}}\right )
$$
and similarly for $W^{(b)}$.
Comparing the singular parts near the branch points,
we obtain the equalities
$$
C^{(b)}_k \, \p_{T^{(a)}} E_k =
C^{(a)}_k \, \p_{T^{(b)}} E_k\,,
\;\;\;\;
C^{(b)}_k \, d W^{(a)}(\lambda_{k}^{*})=
C^{(a)}_k \, d W^{(b)}(\lambda_{k}^{*})
$$
for all $k$.
Dividing one by another, we obtain a closed
system of differential equations of the hydrodynamic type
for the branch points $E_k$:
\beq\label{W3}
\p_{T^{(a)}} E_k =V^{(ab)}_k(E_1 , \ldots \, , E_M)\,
\p_{T^{(b)}} E_k \,,
\;\;\;\; k=1, \ldots , M
\eeq
Here the ``group velocities'' $V^{(ab)}_k$ are given by:
\beq\label{VAB}
V^{(ab)}_k =
\left (\frac{dW^{(a)}(z)}{dW^{(b)}(z)}\right )_{z=E_k}
\eeq
In general they are complicated nonlinear
functions of $E_k$.
The system (\ref{W3})
is diagonal with respect to the $E_k$'s,
so the branch points
of the Schwarz function are Riemann invariants.

Admissible Cauchy data for
the system (\ref{W3}) are given by
the ordinary differential equations of the form
\beq\label{W4}
\p_{T^{(a)}} E_k =
\left (\frac{dW^{(a)}(z)}{dS(z)}\right )_{z=E_k}
\eeq
which is equivalent to the ``string equation''
(\ref{lge1}).
Similarly to the calculation given above,
these equations are obtained by expanding (\ref{lge1})
near zeros of $dE$. Note that $\p_T E(W,T)$ at any
zero $\lambda_{k}^{*}$ of $dE$
is equal to $\p_T E_k$.

Solutions of these equations can be constructed
by means of the
generalized hodograph method \cite{Tsarev,kri1}.
In the version of \cite{kri1}, these solutions
are obtained by the representation of the generating differential
in the form
\beq\label{gen3}
QdE = d\Lambda -T^{(a)} d W^{(a)}- T^{(b)} d W^{(b)}
\eeq
where $d\Lambda$ is a differential with
$T$-independent singularities:
$\p_{T}(\Delta \Lambda (z)) =0$ ($\Delta$ is the
Laplace operator).
For algebraic domains, $d\Lambda$
is a meromorphic differential with fixed
principal parts at all poles.
Since $Q$ is a regular
function on the back side of $\double$
(except for infinity), the r.h.s. of
(\ref{gen3}) must vanish at the points $\lambda_{k}^{*}$.
This requirement leads to the ``hodograph relations''
$$
T^{(a)}dW^{(a)}(E_k )+
T^{(b)}dW^{(b)}(E_k )=
d\Lambda (E_k)
$$
or, equivalently,
\beq\label{hodograph}
T^{(b)}+V_{k}^{(ab)}T^{(a)} = U_{k}^{(b)}
\eeq
for all $k=1, \ldots , M$, where
$$
U_{k}^{(b)}=\left (\frac{d \Lambda (z)}{dW^{(b)}(z)}
\right )_{z=E_k}
$$
They give an implicit solution to the Whitham
equations (\ref{W3}) and to the ``string equation''
(\ref{W4}). The Laplacian growth with a sink at
the point $b$ corresponds to changing $T^{(b)}$
with keeping $T^{(a)}$ fixed.

As a rool, the solutions constructed have bifurcation
points.
In applications to nonlinear waves,
such a point means that the wave front
turns over. In the Laplacian growth, the bifurcation
points correspond to the unphysical
singularities (cusps) of the
moving interface, which are however typical \cite{RMP,SB}
in the limit of zero surface tension.

\section{Example: $N=1$}

Consider an example, where the Schwarz function has a simple
pole at a point $\pole \in \Doil$
and is regular everywhere else in $\Doil$ ($N=1$).
Here we follow refs. \cite{BBB}.
In the vicinity of the pole we have
$$
S(z)=\frac{\mu}{\pole -z} +O(1)\,,
\;\;\;\;\; z\to \pole ,
\;\;\;\; \pole \in \Doil
$$
where the residue is put equal to $-\mu$.
According to the general argument, $S(z)$ for $N=1$
obeys an algebraic equation
$$
\sum_{l,k=0}^{2}A_{lk}S^l(z)z^k =0
$$
which is quadratic in each variable.
Therefore,
the Riemann surface of the Schwarz function
has two sheets over the $z$-plane. One of them
is ``physical''. It is the sheet where the relation
$S(z)=\bar z$ holds on the contour.
In what follows, we call it the first (or upper) sheet.
The pole $\pole$
is on this sheet. The analytic continuation
of the Schwarz function inside water droplets
has another pole at a point
which is the value of the function $E$ at
the simple pole of the function $Q$
at $\infty$ on the back side of the Schottky double.
This pole, $\qole$, lies on the second (unphysical) sheet.
To summarize, the poles of the Schwarz function are
as follows:
\beq\label{ex1}
S(z)=\left \{
\begin{array}{l}
\displaystyle{\frac{\mu}{\pole -z}} +O(1)\,,
\;\;\;\;\; z\to \pole \;\;\mbox{on the 1-st (physical) sheet}
\\ \\
\displaystyle{\frac{\nu}{\qole -z}}+O(1)\,,
\;\;\;\;\; z\to \qole \;\;\mbox{on the 2-nd sheet}
\end{array}
\right.
\eeq
The unitarity condition implies that
\beq\label{ex2}
S(z)=\left \{
\begin{array}{l}
\displaystyle{\bar \qole - \frac{\bar \nu}{z}} + O(z^{-2})\,,
\;\;\;\;\; z\to \infty \;\;\mbox{on the 1-st sheet}
\\ \\
\displaystyle{\bar \pole -\frac{\bar \mu}{z}}+O(z^{-2})\,,
\;\;\;\;\; z\to \infty \;\;\mbox{on the 2-nd sheet}
\end{array}
\right.
\eeq
Therefore, all singularities
of the differential $Sdz$ on the
Riemann surface of the Schwarz function
are two simple poles
at $\pole$ and $\qole$ on different sheets (with residues
$-\mu$, $-\nu$) and two second order poles at two infinities
(with residues $\bar \nu$ and $\bar \mu$).
Sum of the residues must be zero, so $\mu +\nu$
is a real number:
\beq\label{sumres}
\mu +\nu =\bar \mu +\bar \nu
\eeq

Substituting the principal parts of the Schwarz function
(\ref{ex1}) into the equation
\beq\label{ex3}
\tilde z^2 z^2 +b \tilde z^2 z +\bar b \tilde z z^2 +
c \tilde z z
+d\tilde z^2 +\bar d z^2 +e \tilde z +\bar e z +h =0
\eeq
where $\tilde z =S(z)$,
we can fix all the unknown coefficients
except for the free term $h$ (cf. \cite{KM}):
$$
\begin{array}{l}
b=-\pole -\qole
\\ \\
c=|\pole +\qole |^2 +\mu +\nu
\\ \\
d=\pole  \, \qole
\\ \\
e=-\pole \, \qole \, (\bar \pole +\bar \qole)-\pole \nu -\qole \mu
\end{array}
$$
(it is implied that $\mu , \nu \neq 0$).

How many droplets may exist for $N=1$?
Equation (\ref{ex3}) in general defines a smooth curve
of genus $1$ (an elliptic curve).
The real section of this curve (the set of
points such that $\tilde z =\bar z$) consists of at most
two disjoint closed contours on the plane.
If there are two contours, a detailed
analysis shows (see, e.g., the proof
of Theorem 7 in \cite{gus} or the corresponding
example in \cite{BBB})
that one of them necessarily
lies on the unphysical sheet. Therefore, two droplets
are impossible for $N=1$.
In other words, all algebraic domains with $N=1$ are
simply-connected.
In this case the curve (\ref{ex3}) is a rational
(genus-0) curve
with singular points
(a smooth genus-1 curve of the form (\ref{ex3}) can not
be birationally equivalent to the genus-0 Schottky
double of a simply-connected domain).
Given the coefficients $b,c,d,e$, this condition
determines $h$.

Solving the quadratic equation
(\ref{ex3}),
we obtain the Schwarz function in a more explicit form:
\beq\label{ex4}
S(z)=\frac{1}{2}(\bar \pole +\bar \qole)
+\frac{\pole \nu +\qole \mu -(\mu +\nu )z +
(\bar \qole  -\bar \pole )\sqrt{P_4 (z)}}{2(z-\pole )(z-\qole )}
\eeq
Here $P_4(z)$ is a polynomial of 4-th degree
with the leading term $z^4$. Generally
speaking, the function $\sqrt{P_4(z)}$ has four branch points
which are simple roots of the polynomial $P_4$.
The degeneration of the curve occurs when two roots,
say $E_3$ and $E_4$,
merge:
$$
\sqrt{P_4 (z)}=(z-E_3 ) \sqrt{P_2 (z)}\,,
\;\;\;\;\;
P_2 (z)=(z-E_1 )(z-E_2)
$$
so there is only one cut
between $E_1$, $E_2$ and the curve is of genus 0.
The second cut degenerates into the double point of the curve.
Comparing with (\ref{ex2}), one can see that the
1-st (physical) sheet corresponds to the branch of
$\sqrt{P_2(z)}$ such that $\sqrt{P_2(z)} \to z$ as
$z\to \infty$.

Endpoints
of the cut $E_1$, $E_2$
in general can not be found explicitly in a simple form
(in the case at hand one has to solve the equation
of 4-th degree).
We know, however, that they
obey the equations of the hydrodynamic type and
the solutions can be represented in the hodograph
form (\ref{gen3}).

\begin{figure}[tb]
\epsfysize=4cm
\centerline{\epsfbox{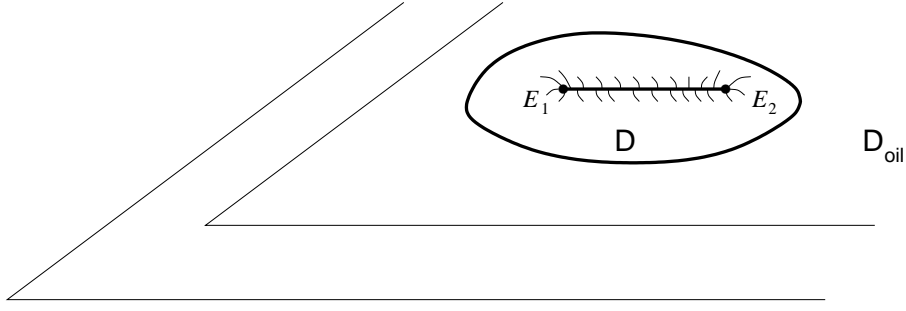}}
\caption{\sl The Riemann surface of the Schwarz function
for $N=1$ is a two-sheeted covering of the $z$-plane.
The map $E: \double \to {\bf P}={\bf C}\cup \{\infty \}$
is a two-sheeted covering, i.e., each point $z\in {\bf P}$
has two pre-images (counting multiplicities).
The image of the front side of the double is the
domain $\Doil$ on the first sheet. The image of the
back side of the double is the rest part of the two-sheeted
Riemann surface. Namely, it is the interior of the
boundary contour on the first sheet (the domain ${\sf D}$)
and the whole second sheet.}
\label{fi:sheets}
\end{figure}

The hodograph relation follows from
the analytic properties of the differential $Sdz$.
It is a meromorphic differential on the 2-sheeted
Riemann surface with a branch cut between
$E_1$, $E_2$. It has
two simple poles and two poles of second order.
The simple poles are $\pole$ on the 1-st sheet
and $\qole$ on the 2-nd sheet with residues
$-\mu$, $-\nu$ respectively.
The second order poles are $\infty^{(1)}$
($\infty$ on the 1-st sheet)
and $\infty^{(2)}$ ($\infty$ on the 2-nd sheet)
with residues $\bar \nu$ and $\bar \mu$ respectively.
Therefore, we can write
\beq\label{sdz}
S dz = \bar \qole dW_{+} +\bar \pole dW_{-}
-\mu dW_{\pole ^{(1)},\, \infty^{(1)}}
+\bar \mu dW_{\infty^{(2)},\, \qole ^{(2)}}
-(\bar \mu -\nu ) dW_{\infty^{(1)},\, \qole ^{(2)}}
\eeq
Here $dW_{\pm}$ are meromorphic differentials
with the only pole of second order at infinity on the
first and second sheets respectively with the principal
part $(1+ O(z^{-2})) dz$ as $z\to \infty$ and
$dW_{a,b}$ is the dipole meromorphic differential
with simple poles at $a$ and $b$ with residues
$\pm 1$ respectively. The upper indices indicate the sheet
where the points lie. For example,
$dW_{\infty^{(1)}, \qole ^{(2)}}$ is the dipole differential
with poles at infinity on the first sheet and $\qole$
on the second sheet.
Explicitly, these differentials
are:
$$
dW_{\pm}=
\left [ \sqrt{P_2(z)}\pm \Bigl ( z-\frac{E_1 +E_2}{2}
\Bigr )
\right ]
\frac{dz}{2\sqrt{P_2(z)}}
$$
$$
dW_{a,b}=
\left ( \frac{\sqrt{P_2(z)}+\sqrt{P_2(a)}}{z-a} -
\frac{\sqrt{P_2(z)}+\sqrt{P_2(b)}}{z-b} \right )
\frac{dz}{2\sqrt{P_2(z)}}
$$

Plugging the explicit formulas for
the differentials into (\ref{sdz}),
we represent $S(z)$ in the form
$$
S(z)=f_1 (z)+\frac{f_2(z)}{\sqrt{P_2(z)}}
$$
where
$$
\begin{array}{l}
\displaystyle{
2f_1(z)=\bar \pole +\bar \qole +
\frac{\mu}{\pole -z}+\frac{\nu}{\qole -z}}
\\ \\
\displaystyle{
2f_2(z)=\frac{1}{2} (\bar \qole -\bar \pole )(2z -E_1 -E_2)
+\bar \mu -\bar \nu +
\mu \, \frac{\sqrt{P_2(\pole )}}{z-\pole }
+\nu \, \frac{\sqrt{P_2(\qole )}}{z-\qole }}
\end{array}
$$
The hodograph equations follow from the fact that
the values $S(E_1)$ and $S(E_2)$
are finite, i.e., $f_2(E_1)=f_2 (E_2)=0$.

If all the parameters are real and $q<E_1 , E_2 <p$,
the system of hodograph equations
acquires the form
\beq\label{hod2}
\left \{
\begin{array}{l}
\displaystyle{
\mu \left (
\sqrt{\frac{\pole -E_2}{\pole -E_1}} -
\sqrt{\frac{E_2 -\qole }{E_1 -\qole }}
\right )
+T\left (\sqrt{\frac{E_2 -\qole }{E_1 -\qole }}-1 \right )=
\frac{1}{2}(E_2 -E_1)
(\pole - \qole )}
\\ \\
\displaystyle{
\mu \left (
\sqrt{\frac{\pole -E_1}{\pole -E_2}} -
\sqrt{\frac{E_1 -\qole }{E_2 -\qole }}
\right )
+T\left (\sqrt{\frac{E_1 -\qole }{E_2 -\qole }}-1 \right )=
-\, \frac{1}{2}(E_2 -E_1)
(\pole - \qole )}
\end{array}
\right.
\eeq
where square roots of positive numbers are assumed
to be positive.
The real parameters $\mu$, $T$, where
$$
T=\mu -\bar \nu =\frac{1}{2\pi i}\oint_{\gamma}
S(z)dz =
\frac{\mbox{Area}\, ({\sf D})}{\pi}
$$
are local coordinates
in the two-dimensional variety of contours obtained from
a circle (of radius $r_0$)
by Laplacian growth processes with oil pumps
located at $p$ and $\infty$. In this interpretation,
$T-r_{0}^{2}$ is the total amount of sucked oil and
$\mu$ is the amount of oil sucked from the point $p$.

\section*{Acknowledgments}

The author is grateful to O.Agam,
E.Bet\-tel\-heim, I.Kri\-che\-ver,
A.Mar\-sha\-kov, M.Mi\-ne\-ev-\-Wein\-stein,
R.Theo\-do\-res\-cu and P.Wieg\-mann
for collaboration and useful discussions.
This work was supported in part by RFBR grant
04-01-00642, by grant for support of scientific schools
NSh-1999.2003.2 and
by the LDRD project 20020006ER ``Unstable
Fluid/Fluid Interfaces" at Los Alamos National Laboratory.

\end{document}